\documentclass[pra,10pt,twocolumn,superscriptaddress,showpacs]{revtex4-1}
\usepackage{amsmath,amsthm}
\usepackage[T1]{fontenc} 
\usepackage{latexsym}
\usepackage{amssymb}
\usepackage{mathtools}
\usepackage{braket}
\usepackage{graphicx}
\usepackage{dsfont}
\usepackage[colorlinks=true, citecolor=blue, urlcolor=blue]{hyperref}
\usepackage{amsfonts}
\makeatletter
\newtheorem*{rep@theorem}{\rep@title}
\newcommand{\newreptheorem}[2]{%
\newenvironment{rep#1}[1]{%
 \def\rep@title{#2 \ref{##1}}%
 \begin{rep@theorem}}%
 {\end{rep@theorem}}}
\makeatother
\usepackage{amsmath}
\usepackage{amssymb}
\usepackage{mathtools}
\usepackage{graphicx}
\usepackage{hyperref}

\usepackage{braket}
\usepackage{dsfont} 
\usepackage{float}
\usepackage{subcaption}
\usepackage{booktabs} 
\usepackage{color} 

\usepackage{txfonts}

\newtheorem{theorem}{Theorem}
\newreptheorem{theorem}{Theorem}

\begin{document}
	\title{Genuine multipartite entanglement detection with mutually unbiased bases (MUBs)}
	

	\author{Sumit Nandi}
		\email{sumit.enandi@gmail.com}
			\affiliation{Purandarpur High School, Purandarpur, West Bengal, 731129, India}
	
	\begin{abstract}
In the present paper, a novel framework to detect genuine multipartite entanglement (GME) has been presented by computing correlations in mutually unbiased bases (MUBs). It has been shown that correlation obtained by measuring in MUBs of all biseparable multipartite states satisfy a bound, whereas GME states violate it. Thus, the presented framework turns out to be a sufficient criterion to detect entanglement in many body scenario. The methodology paves a suitable way to demonstrate certification of different classes of tripartite and quadripartite GME states. In addition to its operational universality, correlation obtained by measuring in MUBs is shown to be juxtaposed with few well known genuine tripartite and quadripartite entanglement measures.
	\end{abstract}
	\maketitle
\section{Introduction}
Quantum Mechanics hosts many counter-intuitive aspects which are not present in the classical description of physical systems. \emph{For example}, uncertainty principle \cite{heisenberg} lays off one of the elemental cornerstones towards the foundation of quantum mechanics. On the other hand, during its early development, it was discovered that quantum theory predicts a nonlocal correlation namely entanglement, between remotely separated components of a given system. Surprisingly, these two mentioned aspects of quantum mechanics are intricately intertwined to each other \emph{i.e.} it may be possible to explain entanglement with the help of complementarity stemming from the uncertainty principle. Notion of complementary basically states that there exist observables that cannot
be measured simultaneously. Mathematically, there exists pairs of observables for which no common eigenbasis can be
found. Consequently, if two observables are complementary to each other, then, it is impossible to prepare a system such that outcomes of measurements in both bases is predictable with certainty - least disturbed outcome in one basis necessarily indicates maximal ignorance in other basis of the pair. The extreme case of
complementarity occurs when the eigenbases of two observables belong to a pair of mutually unbiased bases (MUBs) (see \cite{durt} for a detailed review).
A set of orthonormal bases $B_{1(2)}=\{b_i\}_{i=0}^{d-1}$ are in Hilbert space $C^d$ is called mutually
unbiased if and only if the following holds:
\begin{equation}
 \langle{b_i}|c_j\rangle^2 =
\frac{1}{d}, \forall i,j \in {0, . . . ,d-1}
\end{equation}
A set of orthonormal bases $\{B_1, B_2, \cdots, B_n\}$ in $C_d$ defines a set of mutually unbiased bases if any
two distinct bases belonging this collection are mutually unbiased.
In dimension $d = 2$, a set of three mutually unbiased bases
is readily obtained from the eigenvectors of the three Pauli
matrices $\sigma_x$, $\sigma_y$, $\sigma_z$. The importance of MUBs was first introduced in \cite{ivon} in the context of quantum state determination. In information science, MUBs can be efficiently utilised in quantum state tomography \cite{tomo}, quantum key distribution protocols \cite{qkd}. \\~\\
In a pioneering work \cite{1huber}, MUBs was elegantly used to certify entanglement. The authors presented a formalism to derive powerful entanglement
detection criteria for arbitrarily high-dimensional systems. In a later work, a framework for quantifying entanglement in multipartite and high dimensional systems using only correlations in two MUBs \cite{2huber}. Subsequently, the author \cite{3huber} also proved that measurements only in \emph{two} MUBs are sufficient to detect higher dimensional entanglement. As a follow up, importance of MUBs in certifying entanglement was explored in \cite{4huber}. Moreover, several works can be found in literature which unveils efficient usages of MUBs towards detection of entanglement. The authors \cite{wudarski} provided a class of entanglement witnesses constructed in terms of mutually unbiased bases (MUBs) which reproduces many well known examples such as the celebrated reduction map and the Choi map together with its generalizations. It shows the immense potential of MUBs for detecting entanglement. Along the same direction, the authors in \cite{nature} provided a new family of positive, trace-preserving maps based on mutually unbiased measurements. Entanglement detection using mutually unbiased measurements (MUMs) has been studied in \cite{chen}, thereby, providing a quantum separability criterion that can be experimentally implemented for arbitrary d-dimensional bipartite systems. Along the same line highly symmetric MUMs is investigated to formulate separability criterion in \cite{lai}. The authors in \cite{tavakoli} developed local realist inequalities which is violated maximally by $d$-dimensional MUBs. 
\\
\\
In this paper we take MUB based entanglement detection criterion a step further by developing  a rigorous method to certify genuine multipartite entanglement (GME). Multipartite entanglement is always considered to be a precious resource for carrying out quantum information processing protocols \cite{lee,yeo,gao,epping,liu,sn2020}.
However, detection of multipartite entanglement suffers non trivial technical complexities, and hence, remains largely unexplored, see (\cite{gunhe}) for an extensive review in this regard. Nonetheless, there exists several effective approaches to detect GME \cite{horodecki, hasan,5huber,6huber,sn2024,
bruss,acin1}. A major difficulty in understanding of multipartite entanglement may occur due to the following reason - unlike bipartite entanglement, there merely exists notion of Schmidt decomposition, hence, multipartite entanglement is less canonical
than its bipartite counterpart \cite{acin2}. Naturally, the formalism presented in \cite{1huber} cannot be used in a straightforward manner to detect multipartite entanglement. In the present paper, I generalise the framework of MUB based entanglement detection in multipartite domain, specifically GME states, by computing mutual predictability in \emph{two} different MUBs. It is immediately discovered that the present framework can detect both inequivalent classes of genuine tripartite entangled states, and it can also be generalised in higher dimensional systems. To further explore the potentiality of the approach, we compare MUB based GME detection criterion with some well known measures of multipartite entanglement. \\~\\     
The paper is organised as follows. In Sec. (\ref{bipartite}), we  revisit entanglement detection via MUBs, and discover few of its novel features. In Sec. (\ref{tripartite}), MUBs based GME detection formalism in tripartite scenario is presented. In the next section (\ref{4partite}),  MUBs based GME detection formalism is generalised in quadripartite scenario.  Finally, we note down some concluding remarks in Sec. (\ref{discuss}).  
 
\section{Entanglement detection via MUBs}\label{bipartite}

Underlying symmetry of entangled states implies if a state is non-locally correlated, then measurement outcome of one subsystem in a given basis can be predicted with certain degree of certainty depending on entanglement of the system by knowing outcome of measurement(s) of the another subsystem(s). Correlation as such remains invariant under local unitary transformations which further elucidates if a given state $\ket{\Psi}$ is entangled in certain basis, then it must be entangled in some other basis that is unitarily related to the former one. This very fact has been exploited in \cite{1huber} to investigate separability issue, and the authors inscribed therein a formalism to detect entanglement by employing measurements in two mutually unbiased basis (MUB). Let us consider a correlation function for two quantum observables $\{a, b \in B_1 \}$ which are spanned by orthonormal basis $\{\ket{i_a}\}$($\{\ket{i_b}\}$). We denote the joint probability that the outcome of $a$ is $i$ and the outcome of $b$ is $j$ by $P_{a,b}(i,j)$ given by
\begin{equation}\label{pAB}
P_{a,b}(i,j)=\langle i_a|\otimes\langle j_b|\rho|i_a\rangle\otimes|j_b\rangle,
\end{equation} 
where $\rho=|\Psi\rangle\langle\Psi|$. Then one can construct the correlation function as follow
\begin{equation}\label{cAB}
C_{a,b}=\sum_{i=0}^{d-1}P_{a,b}(i,i),
\end{equation}
where $C_{a,b}(i,j)$ is defined as mutual predictability. It can be used to quantify probability of predicting the measurement outcome of $a$ knowing the outcome of $b$, and \emph{vice versa}. But alone this mutual predictability obtained in certain basis would not be capable of detecting entanglement. For example, $C_{a,b}$, in computational basis, is \emph{one} for both product state $\ket{00}$ and entangled state $\frac{1}{\sqrt 2}(\ket{00}+\ket{11})$. For this reason, correlation functions in \textit{two} \emph{two} MUBs had been considered. Thus, one requires mutual predictability pertaining to the observables  $\{a^\prime, b^\prime\}\in B_2 $ which is mutually unbiased to $B_1$. Define,
\begin{equation}
C_{a^\prime,b^\prime}=\sum_{i=0}^{d-1}P_{a^\prime,b^\prime}(i,i).
\end{equation}  
In the bipartite case, it can be obtained by recasting $\ket{\Psi}$ into Schmidt decomposition form in the corresponding MUBs, and then computing $P_{ab}(i,j)$ using Eq.(\ref{pAB}). Finally, one finds the quantity $I_{2}$ given by 
\begin{equation}
I_{2}=C_{a^\prime,b^\prime}+C_{a,b}
\end{equation}

\begin{figure}
 \includegraphics[width=0.76\linewidth]{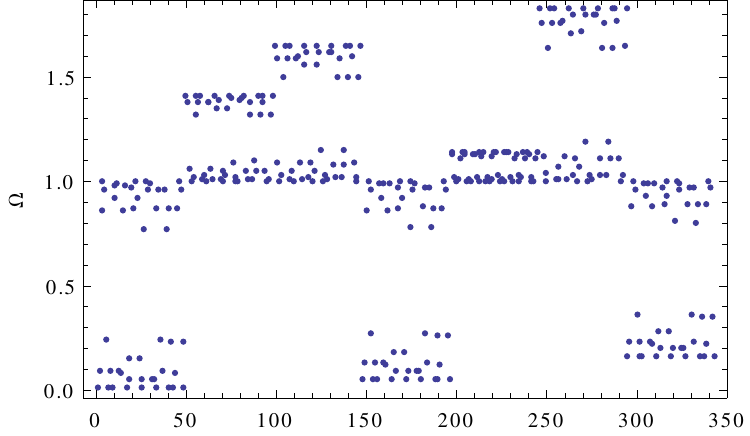}
  \caption{The density plot is obtained by computing the lhs of Eq.(\ref{eq.10}) for the Bell state. Numerically, it shows that the value of the lhs denoted as $\Omega$ always satisfies the bound given by Eq.(\ref{eq.10}) by assuming non-negative values. To obtain the plot, the value of the parameter $\Theta$ is set zero.}\label{locc}
\end{figure}

The authors in \cite{1huber} have presented a separability criterion that relies on a upper bound of $I_{2}$  for separable states. For a pair of MUBs we have $I_2=1+\frac{1}{d}$. For, $m$ MUBs we have $I_m=1+\frac{m-1}{d}$. In particular, for a complete set of MUBs, we have
\begin{equation}\label{eq6}
I_{d+1}=\sum_{k=1}^{d+1}c_{k,k}\le 2.
\end{equation}
Here, we will further investigate some remarkable features of $I_2$ that fully qualifies it to play 
an instrumental role to characterise entanglement.
Consider a general pure state of the form $\ket{\Psi_\lambda}=\sqrt{\lambda}\ket{00}+
\sqrt{1-\lambda}\ket{11}$. It is straightforward to verify the value of $I_2$ of the state given by $\frac{3}{2}+\sqrt{\lambda(a-\lambda)}$. Evidently, $I_2$ violates the bound given in Eq.(\ref{eq6}) for any non-zero value of $\lambda$. Thus, the bound given by Eq.(\ref{eq6}) is necessary and sufficient to certify pure bipartite entanglement. In what follows the set of pure states $\mathcal{S}$ satisfying the inequality   
describes a convex set.\\~\\ 
\textit{Proposition:} The set $\mathcal{S}=\{\rho:I_2(\rho)\le 2\}$ is convex.\\
\textit{Proof:} Let $\omega_1$, $\omega_2$ $\in$ $\mathcal{S}$, and $I_2(\omega_i)\le 2$. Consider $\omega_l=\lambda\omega_1+(1-\lambda)\omega_2$, where $\lambda\in\{0,1\}$. It readily follows that 
\begin{eqnarray}
I_2(\omega_l)&=&\lambda I_2(\omega_1)+(1-\lambda)I_2(\omega_2)\\
&\le & 2\lambda+2(1-\lambda)\\
&\le& 2
\end{eqnarray}
Thus $\rho_l\in \mathcal{S}$, and hence, $\mathcal{S}$ is convex.    \\
\\
Next, we shall explore the characteristic of $I_2$ in local operation and classical communication (LOCC) paradigm. We show that the quantity is non-increasing under LOCC operation. Below, we state one of the central results of the present paper:\\~\\
\emph{Conjecture:} The quantity $I_2$ is monotonically non-increasing under LOCC operations.\\
 We will numerically prove it by following the framework \cite{durpra} for envisaging monotonicity property of $I_2$ under LOCC. Consider whole classes of LOCCs consist of POVMs (positive operator valued measurements) $\{E_1, E_2\}$ acting locally on a particular subsystem of a bipartite density matrix $\rho$, and which produce only binary outcomes $i$ $\in\{0,1\}$. 
Furthermore $E_i$'s satisfy the condition $E_1^\dagger E_1+E_2^\dagger E_2=\mathbb{I}$. Now, we have to show that $I_2$ is non-increasing under the above POVMs \emph{i.e.}
\begin{equation}\label{eq.10}
I_2(\rho)-p_1I_2(\rho_1)-(1-p)I_2(\rho_2)\ge0,
\end{equation} 

%

where $\rho_k=\frac{(E_k\otimes I).\rho.(E_k\otimes I)^\dagger}{p_k}$, $p_k=Tr[\rho (E_k\otimes I)(E_k\otimes I)^\dagger]$. In order to evaluate the above inequality, we write $E_i=\mathcal{D}_i V$ where $\mathcal{D}_b$ ($b=1,2$) and $V$ are given as below
\begin{equation}
\mathcal{D}_1 = \begin{bmatrix} 
	\sin\chi &0 \\[0.2cm]
 0& \sin\zeta\ 
	\end{bmatrix}
\end{equation}
	
\begin{equation}
	\mathcal{D}_2 = \begin{bmatrix} 
	\cos\chi &0 \\[0.2cm]
 0& \cos\zeta\ 
	\end{bmatrix}
\end{equation}

\begin{equation}
	V = \begin{bmatrix} 
	\cos\xi &-e^{i\Theta}\sin\xi \\[0.2cm]
 \sin\xi& e^{i\Theta}\cos\xi\\ 
	\end{bmatrix},
\end{equation}
and the parameters $\chi$, $\zeta$, $\xi$, and $\Theta$ $\in\{-\pi,\pi\}$. We set $\Theta=0$ and numerically verify that LOCC operation indeed satisfy the condition given by Eq.(\ref{eq.10}), see Fig.(\ref{locc}).


\section{Genuine tripartite entanglement detection by measurements in MUBs}\label{tripartite}
Before addressing tripartite entanglement detection, we define a benchmark of separability of a finite dimensional system shared between $N$ individuals: a multipartite state $\ket{\psi}_{bi-sep}$ is considered to be a bi-separable state if it can be recast into the following form 
\begin{equation}
\ket{\psi}_{bi-sep}= \sum_ip_i|\phi_i\rangle\langle \phi_i|\otimes |\chi_i\rangle\langle \chi_i|,
\end{equation}  
where $\ket{\phi_l}$, and $\ket{\chi_{l^\prime}}$ characterises subsystem of $M$ partners and remaining $N-M$ partners, respectively.  By virtue of convexity, set of all bi-separable states constitute a set $\mathcal{S}$. The complement of the set $\mathcal{S}$ of biseparable
states is usually known as genuinely multipartite
entangled (GME) states. The present work is to develop a suitable formalism which can differentiate between $\mathcal{S}$ and $\bar{{\mathcal{S}}}$. In the previous section, we noticed that correlation function of separable states satisfy an inequality, whereas, entangled states do violate the same. In what follows, we evaluate correlation function in MUBs of tripartite state to develop a suitable criterion of GME detection.  \\        
Let us consider the joint probability distributions involving three subsystems, $P_{a,b}(i,j,k)$ and  $P_{a^\prime,b^\prime}(i,j,k)$ pertaining to the outcomes of local measurement settings $a,$ $b$ $\in B_1$ and $a^\prime,$ $b^\prime$ $\in B_2$, where it is noted that $B_i$'s are necessarily MUB to each other. Also, without loss of generality,  one of the joint probabilities, say, $P_{a,b}(i,j,k)$, is taken in natural basis to be \emph{unity}. We need to compute $P_{a^\prime,b^\prime}(i,j,k)$ in another basis. Unfortunately, in this case, we cannot leverage flexibility of Schmidt decomposition framework for the most general kind of tripartite state. However, we can adopt the formalism of recasting an arbitrary tripartite state into the particular decomposition formulated in \cite{acin2} involving minimal number of local bases product states (LBPS). In order to evaluate $P_{a^\prime,b^\prime}(i,j,k)$, we rewrite the given tripartite state in the canonical form involving \emph{least} number of LBPS spanned by an orthonormal basis which is mutually unbiased to the former, in our case it is assumed as computational basis.   
Now recall that the canonical form an arbitrary pure tripartite state involves \emph{five} LBPS. Below we put down \emph{five} inequivalent LBPS \cite{acin3}
\begin{eqnarray}\label{ineq lbps}
\{\ket{\tilde{0}\tilde{0}\tilde{0}},\ket{\tilde{0}\tilde{0}\tilde{1}}, \ket{\tilde{1}\tilde{0}\tilde{0}}, \ket{\tilde{1}\tilde{1}\tilde{0}}, \ket{\tilde{1}\tilde{1}\tilde{1}} \}\\
\{\ket{\tilde{0}\tilde{0}\tilde{0}},\ket{\tilde{0}\tilde{0}\tilde{1}}, \ket{\tilde{0}\tilde{1}\tilde{1}}, \ket{\tilde{1}\tilde{0}\tilde{0}}, \ket{\tilde{1}\tilde{1}\tilde{1}} \}\\
\{\ket{\tilde{0}\tilde{0}\tilde{0}},\ket{\tilde{0}\tilde{1}\tilde{0}}, \ket{\tilde{1}\tilde{0}\tilde{0}}, \ket{\tilde{1}\tilde{0}\tilde{1}}, \ket{\tilde{1}\tilde{1}\tilde{1}} \}\\
\{\ket{\tilde{0}\tilde{0}\tilde{0}},\ket{\tilde{1}\tilde{0}\tilde{0}}, \ket{\tilde{1}\tilde{0}\tilde{1}}, \ket{\tilde{1}\tilde{1}\tilde{0}}, \ket{\tilde{1}\tilde{1}\tilde{1}} \}\label{ineq lbps18}\\
\{\ket{\tilde{0}\tilde{0}\tilde{0}},\ket{\tilde{0}\tilde{1}\tilde{0}}, \ket{\tilde{0}\tilde{1}\tilde{1}}, \ket{\tilde{1}\tilde{1}\tilde{0}}, \ket{\tilde{1}\tilde{1}\tilde{1}} \}\label{ineq lbps19}\\
\{\ket{\tilde{0}\tilde{0}\tilde{0}},\ket{\tilde{0}\tilde{0}\tilde{1}}, \ket{\tilde{1}\tilde{1}\tilde{0}}, \ket{\tilde{1}\tilde{0}\tilde{1}}, \ket{\tilde{1}\tilde{1}\tilde{1}}\label{ineq lbps20} \},
\end{eqnarray}
where $\ket{\tilde{i}}$'s constitute an orthonormal basis, and satisfies $\langle\tilde{i}|i\rangle=\frac{1}{2}$ for $i=\{0,1\}$.
 Let us denote the LBPS as $B_{\alpha}$. Considering $B_{\alpha}$, we introduce correlation function
\begin{equation}\label{cab tri}
C_{a^\prime,b^\prime}=\sum_{B_{\alpha}}
P_{a^\prime,b^\prime}(i,j,k),
\end{equation} 
where $i,j,k\in\{0,1\}$, and assume those values permitted by $B_{\alpha}$. However, $B_{\alpha}$ is not unique, and
different inequivalent LBPS $B_{\alpha}$ involves different product bases. Thus $C_{a^\prime,b^\prime}$ yields different values for different such decompositions. To cope up with the difficulty, we maximize $C_{a^\prime,b^\prime}$ over all $B_{\alpha}$, and introduce 
\begin{equation}\label{cmax}
C^{max}_{a^\prime,b^\prime}=max\sum_{B_{\alpha}}
P_{a^\prime,b^\prime}(i,j,k),
\end{equation} 
Now, we are interested  
to the quantity $I_3$ given by 
\begin{equation}
I_3=C_{a,b}+C^{max}_{a^\prime,b^\prime}.
\end{equation}
Let us evaluate $C^{max}_{a^\prime,b^\prime}$. The sum in the RHS of Eq.(\ref{cmax}) must contain all the \textit{five} LBPS. So we write
\begin{eqnarray}\label{cab_mub}
C^{max}_{a^\prime,b^\prime}=P_{a^\prime, b^\prime}(\tilde{0},\tilde{0},\tilde{0})+P_{a^\prime, b^\prime} (\tilde{0},\tilde{0},\tilde{1})
+P_{a^\prime, b^\prime}(\tilde{0},\tilde{1},\tilde{0})\nonumber\\+P_{a^\prime, b^\prime}(\tilde{1},\tilde{1},\tilde{0})+P_{a^\prime, b^\prime}(\tilde{1},\tilde{1},\tilde{1})
\end{eqnarray} 
We compute $P_{a^\prime, b^\prime}(\tilde{0},\tilde{0},\tilde{0})$ for an arbitrary local product base which is mutually unbiased to the former
\begin{eqnarray}
P_{a^\prime, b^\prime}(\tilde{0},\tilde{0},\tilde{0})&=&|\langle 0|\tilde{0}\rangle |^2|\langle 0|\tilde{0}\rangle|^2
|\langle 0|\tilde{0}\rangle |^2\\
&=&\frac{1}{2}.\frac{1}{2}.\frac{1}{2}=\frac{1}{8}
\end{eqnarray}
Each of these terms coming from the individual product bases puts an upper bound for the local states which include separable and bi-separable tripartite state.
\begin{theorem}\label{th1}
 All separable tripartite states satisfy the following inequality
\begin{equation}\label{eq27}
I_3\le 1+\frac{5}{8}=1.625
\end{equation}
\end{theorem}
 
\textit{proof}: For an arbitrary product state (fully separable), we write the state as $\ket{i}\otimes\ket{j}\otimes\ket{k}$. We can choose $a$, $b$ in such a way that $C^{max}_{a,b}=1$. By choosing appropriate unitary transformation, we can recast $\ket{i}\otimes\ket{j}\otimes\ket{k}$ into $\ket{\tilde{i}}\otimes\ket{\tilde{j}}\otimes\ket{
\tilde{k}}$. Next, we evaluate
\begin{eqnarray}
C^{max}_{a^\prime, b^\prime}(i,j,k)&=&|\langle i|\tilde{0}\rangle |^2|\langle j|\tilde{0}\rangle|^2
|\langle k|\tilde{0}\rangle |^2\nonumber\\
&=&\frac{1}{8}.
\end{eqnarray}
Thus, $I_3=1+\frac{1}{8}<1.625$.
Next, we prove for a general arbitrary tripartite pure state. For this purpose we use the following canonical representation of tripartite state \cite{acin3}
\begin{equation}\label{acin_decompositio}
\ket{{\tilde{\Psi}}}=\lambda_0\ket{000}+
e^{i\phi}\lambda_1\ket{001}+
\lambda_2\ket{010}+
\lambda_3\ket{100}+
\lambda_0\ket{111},
\end{equation} 
where $\lambda_l\in{\mathbb{R^+}}$ satisfying normalisation condition $|\lambda_l|^2=1$, and $0\le\phi\le\pi$. In this case we proceed as follow:
\begin{eqnarray}
C^{max}_{a^\prime, b^\prime}(i^\prime,j^\prime,k^\prime)&=&
 \big(|\langle 0|\tilde{0}\rangle |^2|\langle 0|\tilde{0}\rangle|^2
|\langle 0|\tilde{0}\rangle |^2+ \text{other terms }\big)|\lambda_0|^2+\nonumber\\
&&\big(|\langle 0|\tilde{0}\rangle |^2|\langle 0|\tilde{0}\rangle|^2
|\langle 1|\tilde{0}\rangle |^2+ \text{other terms }\big)|\lambda_1|^2+\nonumber\\
&&\big(|\langle 0|\tilde{0}\rangle |^2|\langle 1|\tilde{0}\rangle|^2
|\langle 0|\tilde{0}\rangle |^2+ \text{other terms }\big)|\lambda_2|^2+\nonumber\\
&&\big(|\langle 1|\tilde{0}\rangle |^2|\langle 0|\tilde{0}\rangle|^2
|\langle 0|\tilde{0}\rangle |^2+ \text{other terms }\big)|\lambda_3|^2+\nonumber\\
&&\big(|\langle 1|\tilde{0}\rangle |^2|\langle 1|\tilde{0}\rangle|^2
|\langle 1|\tilde{0}\rangle |^2+ \text{other terms }\big)|\lambda_4|^2+\label{eq30}\nonumber\\
\\
&=&\frac{5}{8}\sum_\kappa|\lambda^2_{\kappa}|\label{eq31}\\
&=&\frac{5}{8}
\end{eqnarray} 
Where, each term in Eq.(\ref{eq30}) adds $\frac{1}{8}$. This completes the proof of the theorem (\ref{th1}). Some remarkable consequences of the bound given by Eq.(\ref{eq27}) can immediately be followed up. The criterion provided by Eq.(\ref{eq27}) is sufficient to certify pure tripartite entanglement. To prove the sufficiency of our theorem we will use the fact that bi-separability is characterised by rank of reduced density matrices $\rho_m$ ($m=1,2,3$) of the given state \cite{linden}. Any bi-separable tripartite state of the form $\ket{\mu}_1\otimes\ket{\nu}_{23}$, $\ket{\mu^\prime}_2\otimes\ket{\nu^\prime}_{13}$, and $\ket{\mu^{\prime\prime}}_3
\otimes\ket{\nu^{\prime\prime}}_{12}$ has \emph{at least} one rank $1$ reduced density matrix. It is realisable that tripartite bi-separable state of the form mentioned above contains \emph{at most} four LBPS. More precisely $C^{max}_{a^\prime,b^\prime}$ in Eq.(\ref{cab_mub}) would involve only $four$ terms. Thus, $C^{max}_{a^\prime,b^\prime}\le\frac{1}{2}$ and $I_3\le1.625$. 

%

The significance of these results can be explored by discussing adaptability of the present formalism for well known classes of genuinely entangled tripartite states. Next, we evaluate $I_3$ for GHZ state and W state. Let the observables $\{a,b\}$ belong to computational basis, and $\{a^\prime,b^\prime\}\in\{+,-\}$ which is mutually unbiased to former basis. First, we take generalised GHZ state given by 
\begin{eqnarray}
\ket{GHZ}&=&\cos{\theta}\ket{000}+\sin\theta\ket{111}\\
&=&f_1\ket{+++}+
f_2\ket{+--}\nonumber\\
&& +f_3\ket{-+-}+
f_4\ket{--+},
\end{eqnarray}
where $f_i$'s are the appropriate function of the state parameter $\theta$. In the last line we recast GHZ state in $\{+,-\}$ basis to obtain optimal $B_\alpha$. One can check that optimality of $I_3$ is achieved by taking $B_\alpha$ given by Eq.(\ref{ineq lbps18}-\ref{ineq lbps20}). Taking pure state decomposition as given by Eq.({\ref{ineq lbps19}}) we obtain 
\begin{eqnarray}
C^{max}_{a,b}(i,j,k)&=&P_{a,b,c}(0,0,0)+P_{a,b,c}(1,1,1)\hspace{.2 in} \text{and}\\
C^{max}_{a^\prime, b^\prime}(i,j,k)&=&P_{a^\prime, b^\prime, c^\prime}(+,+,+)+P_{a^\prime, b^\prime, c^\prime}(+,-,-)+\nonumber\\
&&
 P_{a^\prime, b^\prime, c^\prime}(-,+,-)+P_{a^\prime, b^\prime, c^\prime}(-,-,+)
\end{eqnarray}    
Finally $I_{3}$ turns out to be
\begin{equation}
I_{3}=\frac{1}{8}(13+\sin{2\theta}).
\end{equation}
Now, to assess non-local aspects of the quantity $I_3$, let us compare it with genuine entanglement measure. Here, we continue our discussion bu considering triangle measure \cite{triangle} as a genuine tripartite entanglement measure. For GHZ state, triangle measure $\tau$ yields the following expression
\begin{equation}
\tau(GHZ)=4 \sqrt{\left(1- \sin ^4\theta- \cos ^4\theta\right)^4}
\end{equation} 
Subsequent plot given by Fig.(\ref{ghz}) provides a comparison of $I_3$ and $\tau$ obtained for GHZ state. It shows that both of these measures of entanglement behaves in a similar way with the state parameter $\theta$ attaining maximal value at $\theta=\frac{\pi}{4}$. Next, we consider generalised W-state which is a class of GME tripartite state: 
\begin{eqnarray}
\ket{W}&=&\cos{\theta}\ket{001}+\cos\alpha\sin\theta
\ket{010}+\sin\alpha\sin\theta\ket{100}\nonumber\\
&=&\omega_1\ket{+++}+
\omega_2\ket{+-+}+\omega_3\ket{-++}+
\omega_4\ket{-+-}\nonumber\\&&+\omega_5\ket{---},\nonumber\\
\end{eqnarray} 
 where $\omega_i$'s are suitable function of $\theta$ and 
$\alpha$. In a similar vein, one can obtain $I_3$ for $\ket{W}$ state
\begin{equation}
I_{3}=\frac{1}{8} \left(13-2 \sin \alpha \cos \alpha  \sin ^2\theta+\sin 2 \theta (3 \sin \alpha+\cos \alpha )\right).
\end{equation}
As before, we obtain triangle measure of $W$ state $\tau(W)$ and plot $I_3$, and $\tau(W)$ in Fig.(\ref{wst}) for $\alpha=\frac{\pi}{4}$. It further shows both of these quantities are in juxtaposition with each other. Here we note an important observation. The plot Fig.(\ref{wst}) shows that $I_3$ does not always violate the bound given by Eq.(\ref{eq27}). For the given value of the parameters $\alpha=\frac{\pi}{4}$, and $\theta\in\{1.4,\frac{\pi}{4}\}$, the quantity $I_3$ does not violate the bound, although the state is genuinely entangled. It signifies the fact that the theorem (\ref{th1}) is not a necessary condition to certify GME, nevertheless, it is already shown to be a sufficient criterion.    
\begin{figure}
 \includegraphics[width=1.0\linewidth]{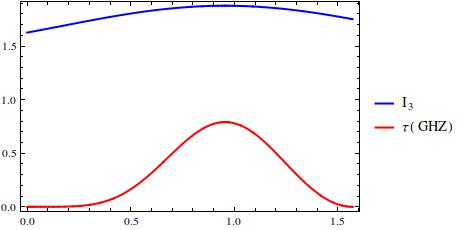}
 \caption{Comparative behaviour of the quantities $I_{3}$ and triangle measure $\tau$ of tripartite generalised GHZ state is shown for $\theta\in\{0,\frac{\pi}{2}\}$. }
 \end{figure}\label{ghz}
 \begin{figure}
  \includegraphics[width=1.0\linewidth]{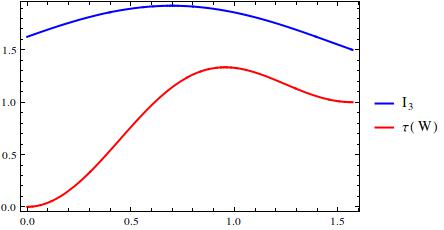}
\caption{Comparative behaviour of the quantities $I_{3}$ and triangle measure $\tau$ of tripartite generalised W state is shown for $\theta\in\{0,\frac{\pi}{2}\}$. To obtain this plot we take $\alpha=\frac{\pi}{4}$.
}
\end{figure}\label{wst}

%
%

\section{Genuine quadripartite entanglement detection by measurements in MUBs}\label{4partite}
Next, we consider the joint probability involving four subsystems, $P_{a,b}(i,j,k,l)$ pertaining to the outcomes of all local measurement settings $a,$ $b$ of a quadripartite state.
The joint probability involving four subsystems, $P_{a,b}(i,j,k,l)$ pertaining to the outcomes of all local measurement settings $a,$ $b$ belongs to a $2^4$ dimensional Hilbert space. However, we consider the following minimum number of LBPS required to specify a four-qubit system as given by \cite{acin3}
\begin{eqnarray}\label{ineq 4lbps}
\{\ket{0000},\ket{0100}, \ket{0101}, \ket{0110}, \ket{1000},
\ket{1001},\nonumber\\ \ket{1010}, \ket{1011}, \ket{1100}, \ket{1101}, \ket{1110},\ket{1111}\}.
\end{eqnarray} 
By choosing appropriate MUBs for each subsystem, one can find $I_4$ \begin{equation}
I_4=C^{max}_{a,b}+C^{max}_{a^\prime,b^\prime},
\end{equation} 
where mutual predictability  $C^{max}_{a^\prime,b^\prime}$ is obtained  as follow 
\begin{eqnarray}
C^{max}_{a^\prime,b^\prime}(\tilde{i},\tilde{j},\tilde{k},\tilde{l})=P_{a^\prime, b^\prime}(\tilde{0}\tilde{0}\tilde{0}\tilde{0})+P_{a^\prime, b^\prime} (\tilde{0}\tilde{1}\tilde{0}\tilde{0})
+P_{a^\prime, b^\prime}(\tilde{0}\tilde{1}\tilde{0}\tilde{1})\nonumber\\+P_{a^\prime, b^\prime}(\tilde{0}\tilde{1}\tilde{1}\tilde{0})+P_{a^\prime, b^\prime}(\tilde{1}\tilde{0}\tilde{0}\tilde{0})++P_{a^\prime, b^\prime}(\tilde{1}\tilde{0}\tilde{0}\tilde{1})\nonumber\\
+P_{a^\prime, b^\prime}(\tilde{1}\tilde{0}\tilde{1}\tilde{0})+P_{a^\prime, b^\prime}(\tilde{1}\tilde{0}\tilde{1}\tilde{1})+P_{a^\prime, b^\prime}(\tilde{1}\tilde{1}\tilde{0}\tilde{0})\nonumber\\+P_{a^\prime, b^\prime}(\tilde{1}\tilde{1}\tilde{0}\tilde{1})+
P_{a^\prime, b^\prime}(\tilde{1}\tilde{1}\tilde{1}\tilde{0})+P_{a^\prime, b^\prime}(\tilde{1}\tilde{1}\tilde{1}\tilde{1})\nonumber\\
\end{eqnarray} 
Here $\ket{\tilde{i}}$ constitutes a basis which is mutually unbiased to the previous one. For example, we can assume $\ket{i}$, and $\ket{\tilde{i}}$ to be computational, and Hadamard basis respectively.

As before $P_{a^\prime, b^\prime}(\tilde{i},\tilde{j},\tilde{k},\tilde{l})$ is generalised in the following way  
\begin{eqnarray}
P_{a^\prime, b^\prime}(\tilde{i},\tilde{j},\tilde{k},\tilde{l})&=&|\langle i|\tilde{i}\rangle |^2|\langle j|\tilde{j}\rangle|^2
|\langle k\tilde{k}\rangle |^2|\langle l|\tilde{l}\rangle|^2
\\
&=&\frac{1}{2^4}
\end{eqnarray}
In quadripartite case, we require \emph{twelve} LBPS to recast an arbitrary state. Hence, each of those terms make contribution, and adding those individuals we obtain an upper bound of $I_4$ which separable states must satisfy. It prompts us to present the following theorem
\begin{theorem}
All separable quadripartite states satisfy the following inequality
\begin{equation}\label{eq27}
I_4\le 1+\frac{3}{4}=1.75
\end{equation}
\end{theorem}
The proof can be generalised as outlined for tripartite case in the previous section. It is noted that our GME criterion of quadripartite entanglement is sufficient to certify genuine entanglement. Any biseparable state must satisfy the bound but the converse does not hold, \emph{i.e.} the criterion is sufficient but not necessary. To illustrate the result, we compare the quantity $I_4$ with a quadripartite entanglement measure for generalised GHZ and generalised W state. In order to compute $I_4$ for GHZ state, we use computational basis $\{\ket{0},\ket{1}\}$ and Hadamard basis $\{\ket{+},\ket{-}\}$ as two representatives of MUBs. Next, we obtain the expressions of mutual predictability in these bases 
 \begin{eqnarray}
C^{max}_{a,b}(i,j,k,l)&=&P_{a, b}(0000)+P_{a, b} (0100)
+P_{a,b}(0101)+\nonumber\\&&P_{a, b}(0110)+P_{a, b}(1000)+P_{a, b}(1001)+\nonumber\\&
&P_{a, b}(1010)+P_{a, b}(1011)+P_{a, b}(1100)+\nonumber\\&&P_{a, b}(1101)+
P_{a, b}(1110)+P_{a, b}(1111) \hspace{.5 in} \nonumber\\
\end{eqnarray}
\begin{eqnarray}
C^{max}_{a^\prime, b^\prime}(i,j,k,l)&=&P_{a, b}(++++)+P_{a, b} (+-++)
+P_{a,b}(+-+-)+\nonumber\\&&P_{a, b}(+--+)+P_{a, b}(-+++)+P_{a, b}(-++-)+\nonumber\\
&&P_{a, b}(-+-+)+P_{a, b}(-+--)+P_{a, b}(--++)+\nonumber\\&&P_{a, b}(--+-)+
P_{a, b}(---+)+P_{a, b}(----) \nonumber\\
\end{eqnarray} 
In the last equation $C^{max}_{a^\prime, b^\prime}$ is obtained by rewriting GHZ state in the Hadamard basis. Thus we obtain the following value of the quantity $I_4$ for GHZ state 
\begin{equation}
I_4=\frac{25+7\sin\theta}{16}
\end{equation}
In order to compare the value of $I_4$ of generalised GHZ state with a genuine entanglement measure, we consider global measure of entanglement denoted as $Q$ \cite{brenen03}, and given by 
\begin{equation}
Q(\ket{\psi})=2\Big(1-\frac{1}{n}\sum_{k=0}^{n-1}
Tr\rho_k^2\Big),
\end{equation}
 where $\ket{\psi}\in C^2\otimes C^2 \otimes C^2 \otimes C^2$, and $\rho_k$'s are the single  qubit reduced density matrix. The global measure for generalised GHZ state yields following expression 
\begin{equation}
Q=\sin^22\theta
\end{equation} 
We plot the quantities $I_4$ and $Q$ in Fig(\ref{4partiteplot}) to visualise a trade off between them. As before, in the whole parameter range of $\theta$, the plots depict identical nature of entanglement viewed from two entirely different frameworks of quantifying entanglement. Both the values of $I_4$ and $Q$ attain maximum at $\theta=\frac{\pi}{4}$ \emph{i.e.} for GHZ state. It is emphasised that in the regions $\theta\in \{0,0.22\}$ and $\theta\in\{1.35,1.57\}$ the state satisfies the bound which is showing that the presented criterion is sufficient but not necessary.
\begin{figure}
 \includegraphics[width=1.0\linewidth]{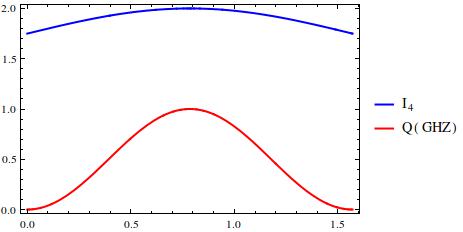}
\caption{Comparative behaviour of the quantities $I_{4}$ and global measure $Q$ of quadripartite generalised GHZ state is shown for $\theta\in\{0,\frac{\pi}{2}\}$.}
\end{figure}\label{4partiteplot}
To complete the discussion, we consider quadripartite generalised W state,
\begin{eqnarray}
\ket{W_g}=\cos{\theta}\ket{0001}+\sin\mu\sin\theta
\ket{0010}\nonumber\\+\cos\mu\sin\nu\sin\theta\ket{0100}
+\sin\mu\sin\nu\sin\theta\ket{1000},\nonumber\\
\end{eqnarray} 
 where $\theta\in\{0,\pi\}$, $\mu\in\{0,\pi\}$
, and $\nu=\frac{\pi}{4}$. It is to be noted that for $\theta \approx 1.04$, $\mu \approx0.62$ and $\nu=\frac{\pi}{4}$ the state $\ket{W_g}$ yields W-state. We study comparative behaviour of $I_4$ and $Q$ in Fig.(\ref{4partiteplotw}) for a specific choice of the parameters given by $\theta=1.05$ and $\nu=0.5$. It shows that the two plots are juxtaposed as is shown earlier. 

\begin{figure}
 \includegraphics[width=1.0\linewidth]{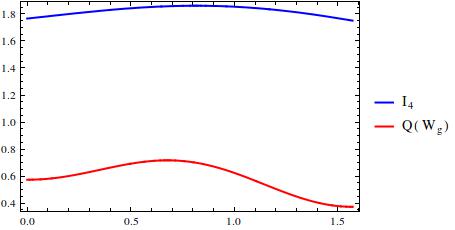}
\caption{Comparative behaviour of the quantities $I_{4}$ and global measure $Q$ of quadripartite generalised W state is shown for $\mu\in\{0,\frac{\pi}{2}\}$. To obtain the plot, we take $\theta=1.05$, and $\nu=0.5$.}
\end{figure}\label{4partiteplotw}
\section{Conclusion}\label{discuss}
Certification of entanglement is a very crucial task in quantum information theory. In this direction, MUB based separability criterion is a very effective tool. We have shown that such correlation obtained for bipartite states is convex, and non-increasing under LOCC. Then, we provide a suitable and rigorous framework of entanglement detection, specifically, genuine multipartite entanglement (GME) detection. The criterion of GME certification inscribed in this work is based on correlation obtained by measuring in mutually unbiased bases. In order to evaluate such correlation for multipartite states, we choose computational basis, and a particular decomposition of the given state that essentially involves minimal
number of LBPS, respectively. Correlation, thus obtained, is satisfied by all biseparable tripartite and quadripartite states. Thus, it provides a sufficient condition to certify tripartite and quadripartite genuine entanglement. However, a general formulation of decomposing an arbitrary multipartite state into LBPS is still an open question. For this reason, the underlined framework might not be generalised for arbitrary multipartite states. Nevertheless, our formulation is very much likely to be accessed experimentally through measuring correlations with fewer measurement settings on the subsystems. We note that a full state tomography requires substantial experimental
effort that grows exponentially with the number of subsystems. Thus, our
approach can be leveraged with much less rigorous effort to detect GME using a 
number of measurement settings that only grows linearly with the number of subsystems.  


%
%

\end{document}